\documentclass{egpubl}
\usepackage{VMV2024}
 
\WsSubmission      

\usepackage[T1]{fontenc}
\usepackage{dfadobe}  

\usepackage{cite}  
\BibtexOrBiblatex
\electronicVersion
\PrintedOrElectronic
\ifpdf \usepackage[pdftex]{graphicx} \pdfcompresslevel=9
\else \usepackage[dvips]{graphicx} \fi

\usepackage{egweblnk} 

\usepackage[printonlyused, nolist]{acronym}
\graphicspath{{figs/}}
\usepackage{tikz}
\usepackage{placeins}
\usepackage{siunitx}
\usetikzlibrary{positioning}
\usepackage{placeins}
\usepackage{tabularray}
\usepackage{enumitem}
\usepackage{subcaption}
\usepackage{multirow}
\usepackage{mathtools}
\usepackage{amssymb}
\usepackage{makecell}
\usepackage{array}

\UseTblrLibrary{booktabs}

\newcommand{\rb}[1]{\left(#1\right)}

\newcommand{\cb}[1]{\left\{#1\right\}}

\newcommand{\hide}[1]{\iffalse #1 \fi}

\newcommand{\nint}{\mathbb{Z}}

\newcommand{\nreal}{\mathbb{R}}

\newcommand{\nrealop}{\mathbb{R}_{\geq 0}}
\newcommand{\nrealp}{\mathbb{R}_{>0}}

\DeclareOldFontCommand{\rm}{\normalfont\rmfamily}{\mathrm}
\DeclareOldFontCommand{\sf}{\normalfont\sffamily}{\mathsf}
\DeclareOldFontCommand{\tt}{\normalfont\ttfamily}{\mathtt}
\DeclareOldFontCommand{\bf}{\normalfont\bfseries}{\mathbf}
\DeclareOldFontCommand{\it}{\normalfont\itshape}{\mathit}
\DeclareOldFontCommand{\sl}{\normalfont\slshape}{\@nomath\sl}
\DeclareOldFontCommand{\sc}{\normalfont\scshape}{\@nomath\sc}

\begin{document}

\title[Towards Practical Meshlet Compression]{Towards Practical Meshlet Compression}

\ifx\anonymize\undefined
	\author[Kuth et al.]
	{
		\parbox{\textwidth}
		{
			\centering 
			Bastian Kuth$^1$\orcid{0000-0001-9473-8847}             
			Max Oberberger$^2$\orcid{0000-0001-9648-3171}
      		Felix Kawala$^{1,2}$
      		Sander Reitter$^{1,2}$
      		Sebastian Michel$^{1}$
			Matth\"aus Chajdas$^2$\orcid{0000-0003-4689-2932}
			Quirin Meyer$^1$\orcid{0000-0001-7073-442X}
			{\parbox{\textwidth}{
			\centering $^1$Coburg University of Applied Sciences and Arts, Germany $\quad ^2$AMD, Germany}}			
		}    
	}
\else
	\author[paper1012]{
		\parbox{\textwidth}
		{
			\centering 
			\vspace{0.5cm}
			paper1012
		}  
	}
\fi

\begin{acronym}[CMAcronyms]
\acro{GPU}{graphics processing unit}
\acro{CPU}{central processing unit}
\acro{ALU}{arithmetic and logical unit}
\acro{API}{application programming interface}
\acro{3D}{three-dimensional}
\acro{2D}{two-dimensional}
\acro{1D}{one-dimensional}
\acro{bpt}{bits per triangle}
\acro{bpv}{bits per vertex}
\acro{LUT}{lookup table}
\acro{SIMD}{single instruction, multiple data}
\acro{AABB}{axis-aligned bounding box}
\acro{GLSL}{OpenGL Shading Language}
\acro{rANS}{range asymmetrical numeral systems}
\acro{POT}{power-of-two}
\acro{LBS}{linear blend skinning}
\acro{DQS}{dual quaternion skinning}
\acro{LMS}{log-matrix skinning}
\acro{SBS}{spherical blend skinning}
\acro{CRS}{centre of rotation-optimized skinning}
\acro{i.i.d.}{independent and identically distributed}
\acro{OSS}{Optimal Simplex Sampling}
\acro{LP}{linear program}
\acro{ILP}{integer linear program}
\acro{MILP}{mixed integer linear program}
\acro{GTS}{generalized triangle strip}
\acro{ATS}{alternating triangle strip}
\acro{CLZ}{count leading zeros}
\acro{TEA}{Tiny Encryption Algorithm}
\acro{ETA}{Enhanced Tunneling Algorithm}
\acro{LW}{Laced Wires}
\acro{Gtps}{Giga triangles per second}
\end{acronym}

\teaser{
    \fontsize{9pt}{8pt}\selectfont%
    \def\svgwidth{\textwidth}%
\begingroup%
  \makeatletter%
  \providecommand\color[2][]{%
    \errmessage{(Inkscape) Color is used for the text in Inkscape, but the package 'color.sty' is not loaded}%
    \renewcommand\color[2][]{}%
  }%
  \providecommand\transparent[1]{%
    \errmessage{(Inkscape) Transparency is used (non-zero) for the text in Inkscape, but the package 'transparent.sty' is not loaded}%
    \renewcommand\transparent[1]{}%
  }%
  \providecommand\rotatebox[2]{#2}%
  \newcommand*\fsize{\dimexpr\f@size pt\relax}%
  \newcommand*\lineheight[1]{\fontsize{\fsize}{#1\fsize}\selectfont}%
  \ifx\svgwidth\undefined%
    \setlength{\unitlength}{610.32796904bp}%
    \ifx\svgscale\undefined%
      \relax%
    \else%
      \setlength{\unitlength}{\unitlength * \real{\svgscale}}%
    \fi%
  \else%
    \setlength{\unitlength}{\svgwidth}%
  \fi%
  \global\let\svgwidth\undefined%
  \global\let\svgscale\undefined%
  \makeatother%
  \begin{picture}(1,0.50116065)%
    \lineheight{1}%
    \setlength\tabcolsep{0pt}%
    \put(0.88244797,0.00306732){\color[rgb]{0,0,0}\makebox(0,0)[t]{\lineheight{1.25}\smash{\begin{tabular}[t]{c}\textbf{(d)} GTS-Reuse: 5.9 bpt\end{tabular}}}}%
    \put(0,0){\includegraphics[width=\unitlength,page=1]{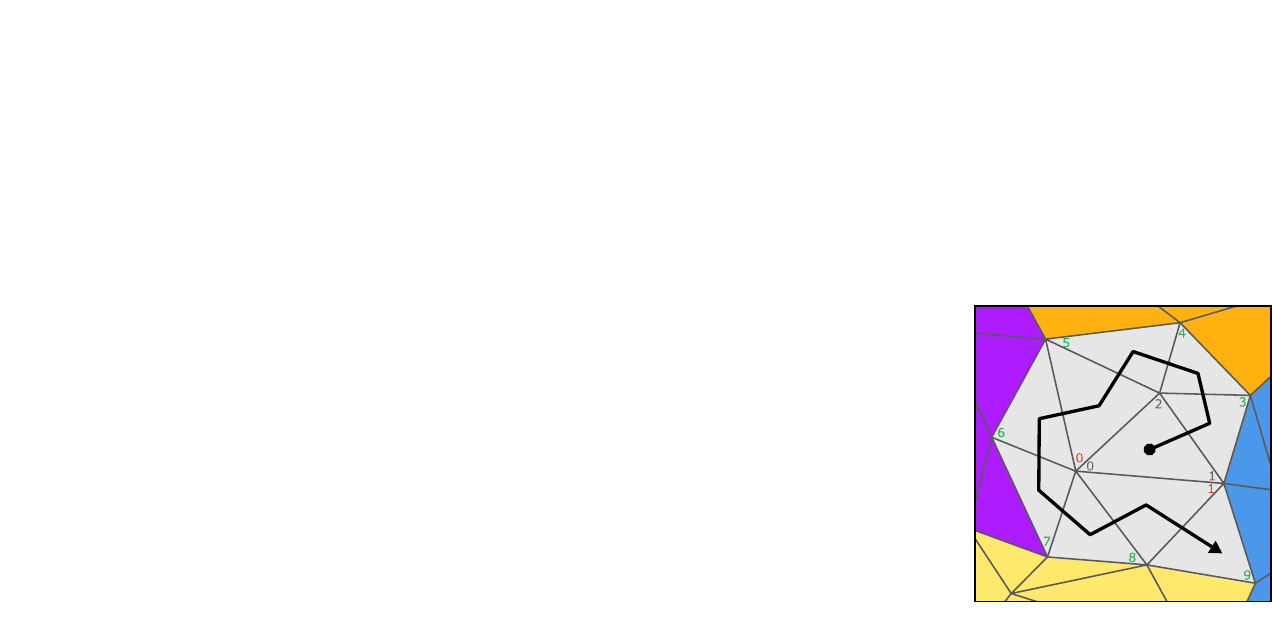}}%
    \put(0.62700251,0.00306732){\color[rgb]{0,0,0}\makebox(0,0)[t]{\lineheight{1.25}\smash{\begin{tabular}[t]{c}\textbf{(c)} GTS: 9.5 bpt\end{tabular}}}}%
    \put(0,0){\includegraphics[width=\unitlength,page=2]{Compression_tex.pdf}}%
    \put(0.37155655,0.00306732){\color[rgb]{0,0,0}\makebox(0,0)[t]{\lineheight{1.25}\smash{\begin{tabular}[t]{c}\textbf{(b) }Basic: 24 bpt\end{tabular}}}}%
    \put(0,0){\includegraphics[width=\unitlength,page=3]{Compression_tex.pdf}}%
    \put(0.11611158,0.00306732){\color[rgb]{0,0,0}\makebox(0,0)[t]{\lineheight{1.25}\smash{\begin{tabular}[t]{c}\textbf{(a) }Vertex: 96 bpt\end{tabular}}}}%
    \put(0,0){\includegraphics[width=\unitlength,page=4]{Compression_tex.pdf}}%
  \end{picture}%
\endgroup%
    \definecolor{teaserGreen}{RGB}{40, 173, 92}%
    \definecolor{teaserRed}{RGB}{235, 59, 35}%
    \centering
    \caption{%
    \textbf{(a)} The conventional vertex pipeline requires $3\times32$ \ac{bpt}, but,
    \textbf{(b)} meshlet-triangles only require $3\!\times\!8$ \ac{bpt}.
    \textbf{(c)} We compress mesh triangles to 9.5 \ac{bpt} by encoding meshlets into optimal \acfp{GTS}.
    \textbf{(d)} We re-order vertices such that their first appearance needs only one bit ({\color{teaserGreen}green}) and only store reused vertex indices ({\color{teaserRed}red}) explicitly.
    }
    \label{fig:teaser}
}

\maketitle

\begin{abstract}
	We propose a codec specifically designed for meshlet compression, optimized for rapid data-parallel \acs{GPU} decompression within a mesh shader.
Our compression strategy orders triangles in optimal \acfp{GTS}, which we generate by formulating the creation as a \acf{MILP}.
Our method achieves index buffer compression rates of 16:1 compared to the vertex pipeline and crack-free vertex attribute quantization based on user preference.
The $15.5$ million triangles of our teaser image decompress and render in 0.59 ms on an AMD Radeon RX 7900 XTX.
\end{abstract}



\section{Introduction}
\emph{Mesh shaders} are a recent addition to the set of programmable shaders used for rendering.
They replace the vertex, tessellation, and geometry stages of the \emph{vertex shader pipeline}.
The resemblance to a compute shader allows for greater flexibility.
With meshes becoming more complex, compression methods keep their memory footprint tangible.
As decompression is usually a slow and serial process, specialized techniques are required for massively parallelized hardware like \acp{GPU}.
In this work, we adapt and extend these techniques to the context of a mesh shader.
We make the following \emph{contributions}:
\begin{itemize}[leftmargin=*]
\item\emph{Meshlet topology compression schemes.}
We reduce the meshlet index buffer memory footprint to ca.\ 5.9 \ac{bpt}.
We propose a slightly faster compression scheme at ca.\ 9.5 \ac{bpt}.
\item\emph{Mesh shader decompression.}
We propose real-time mesh-shader-based decompressing algorithms that outperform the conventional vertex pipeline.
\item\emph{Self-contained, crack-free vertex attribute compression.}
By duplicating meshlet boundary vertices, we keep our meshlets self-contained which allows for fast, spatially-local memory access.
We show a crack-free quantization for attribute compression.
\item\emph{Optimal \ac{GTS} construction.}
Finding the optimal \ac{GTS} is NP-complete.
Nonetheless, we show that it is feasible to find the optimum for common meshlet sizes using a \ac{MILP} solver.
\end{itemize}
Our goal is to provide a mesh-shader-based meshlet decompression technique that is as least as fast the vertex-pipeline while significantly reducing the \ac{GPU} memory footprint.
Our approach has the following \emph{limitations}:
We only account for triangle rasterization and omit other mesh types and render methods.
Due to our tight performance budget, we make limited use of attribute coherence in a meshlet.
Finally, methods for optimally compressible meshlet generation remain future work.

\section{Background and Previous Work}\label{sec:PreviousWork}
A vertex is an \emph{index} $V_i$ referencing an element in a \emph{vertex buffer}.
Vertex-buffer elements contain \emph{vertex attributes} like positions, normals, texture coordinates, etc.
An \emph{index buffer} is a sequence of vertex indices $V_i$ that describe the connectivity.
For \emph{triangle lists}, triplets of indices from a triangle. They have a relatively high memory requirement, but are supported by \acp{GPU}.
 
\emph{Triangle strips} describe a path over a mesh.
A path enters a triangle $(V_a, V_b, V_c)$ through the edge $(V_a, V_b)$.
The next triangle can either be across the \emph{right edge} $(V_b, V_c)$ or the \emph{left edge} $(V_c, V_a)$.
For \emph{\acp{ATS}}, the path always alternates between right and left edge.
Then, three consecutive indices in an index buffer form a triangle.
Except for the first one, each triangle requires only one new index.
\emph{\Acfp{GTS}} are more flexible: for each triangle, one so-called \emph{L/R-flag} per triangle discriminates whether the path continues across the left or right edge~\cite{Velho99HGT}.
Fig.~\ref{fig:teaser} shows an example of a \ac{GTS}.

While \acp{ATS} result in long triangle bands, \acp{GTS} avoid them and allow layouts that better fit \ac{GPU} run-time behavior~\cite[Sec.~7.2]{Kilgard08MOU}.
Further, typically \acp{GTS} require fewer strips than \acp{ATS} and have a smaller memory footprint.
Since \acp{GPU} do not support \acp{GTS} directly, we build upon the algorithm proposed by Meyer et al.~\shortcite{Meyer12DPD} to decode them into triangle lists.

Geometry compression handles the massive and increasing amount of data in real-time computer graphics.
Several overview reports cover this vast field~\cite{Alliez03RAC,Peng05T3M,Maglo153MC}.
While many methods target offline decompression, we narrow our scope to mesh representations suitable for \ac{GPU} rendering.

Calver~\shortcite{Calver2002VDS} describes vertex shader attribute de-quantization using 8- and 16-bit integers for each attribute channel.
Purnomo et al.~\shortcite{Purnomo05HCV} assign a fixed bit-budget for all vertex attributes.
Every attribute channel is quantized with an arbitrary number of bits using a pre-process that compares the rendering error for different bit-allocations.
Kwan et al.~\shortcite{Kwan18PVD} store vertex attributes as 2D block-compressed textures.

For positions, quantization levels are typically determined empirically~\cite{Deering95GC}, usually at 8 -- 12 bits per component~\cite{Peng05T3M, Alliez03RAC}.
To compactly represent positions, Lee et al.~\shortcite{Lee10C3M} first align a mesh on a global grid to prevent cracks.
Next, they decompose a mesh until each sub-mesh does not require more than 8 bits along each spatial axis.
Meyer et al.~\shortcite{Meyer11ALP} dynamically add and remove bits view-dependently. 

For normals, various compression methods exist:
Deering~\shortcite{Deering95GC} proposed spherical parameterization methods, which, however, use expensive trigonometric functions.
Meyer~\shortcite{Meyer12RTG} carefully analyzes the error of various parameterization schemes.
Octahedron unit vectors turn out to be an effective method for vertex-shader decompression~\cite{Meyer10OFN}.
Cigolle and colleagues~\shortcite{Cigolle2014SER} provide effective implementations.
Keinert et al.~\shortcite{Keinert15SFM} propose a fast and precise unit vector decompression method based on a Fibonacci mapping.

Frey and Herzeg~\shortcite{Frey11SSD} convert three perpendicular tangent-space unit-vectors used as vertex attributes required for normal mapping to a quaternion.
They decompress the corresponding tangent-space in a vertex shader.
Recently, methods for blend-attributes compression were presented~\cite{Kuth21VBA, Peters22PCV}.

In his pioneering work, Deering~\shortcite{Deering95GC} introduced real-time index buffer compression.
Offline methods like Edgebreaker~\cite{Rossignac99ECC} or the Cut-Border Machine~\cite{Gumhold98CBM} compress close to the minimum of ca.~1.62 \ac{bpt}~\cite{Tutte62CPT}.
Jakob et al.~\shortcite{Jakob17Pac} provide a fast, but non-real-time, \ac{GPU}-implementation of the Cut-Border Machine.
Meyer et al.~\shortcite{Meyer12DPD} describe a real-time algorithm to quickly decompress a \ac{GTS} using data-parallel scans. 
Karis et. al~\shortcite{karis2021nanite} replace these scans by faster bit-scans, but overall, do not consider their approach to be fast enough for frame-by-frame decompression.

Schäfer et al.~\shortcite{Schaefer12MAT} simplify a mesh with edge-collapses.
The lost geometric information is re-sampled and compactly stored.
During run-time it is reconstructed using hardware-tessellation patterns.
Maggiordomo and coworkers~\shortcite{Maggiordomo23MMC} use a lossy algorithm to convert a highly detailed mesh into the recently introduced micro-mesh~\cite{Nvidia23MM} data structure.
Both approaches significantly reduce memory and rendering time, but change topology, which might not be acceptable for all applications.

Laced Ring~\cite{Gurung11LRC} is a compact data structure for triangle mesh adjacency information.
It targets mesh processing algorithms that require finding triangle neighbors or fans around a vertex.
Hence, it bears some inherent overhead when only used for mesh compression.
However, Mlakar et al.~\shortcite{Mlakar24EEC} showed that is also suited for meshlet decompression and can achieve real-time rates.

\section{Mesh-Shader}\label{Sec:MeshShader}
\newcommand{\Vmax}{\tilde{V}}%
\newcommand{\Tmax}{\tilde{T}}%
\newcommand{\Vcur}{V}%
\newcommand{\T}{T}%
\emph{Amplification-} and \emph{mesh-shaders} provide a compute-shader-based programming model to hand triangles to the raster stage.
Previously, the \emph{vertex pipeline} handled this task.
From the \ac{CPU}, the programmer launches multiple mesh-shader \emph{thread groups} of up to 128 threads.
Hardware internally schedules the thread groups into \emph{waves} of 32 or 64 threads, which then run on a \emph{compute unit}.
Each shader thread-group outputs a small mesh -- called \emph{meshlet} -- to the raster stage.
A meshlet consists of a triangle list stored in a local vertex-buffer with per-vertex attributes and an index-buffer whose elements point into the local vertex-buffer.
Each meshlet has a limit $\Vmax$ vertices and $\Tmax$ triangles, currently set to 256 in DirectX~\cite{DirectXSpecs}.
An \emph{amplification shader} stage \emph{may} run before the mesh-shader stage.
It has the ability to dispatch or suspend mesh-shader thread groups.
We use the amplification shader for a \emph{cone culling} optimization: the triangle normals of a meshlet form a cone that we store using an axis and an opening angle.
Then, the amplification shader can quickly assess, if all triangles are back-facing and then cull the entire meshlet.

Each mesh-shader work-group needs to read in a meshlet from \ac{GPU} memory.
To obtain meshlets in a pre-process, a meshlet split algorithm tries to find coherent groups of triangles such that the total number of vertices $\Vcur$ and triangle $T$ per such group, is smaller than or equal to the given maximum $\Vmax$ and $\Tmax$.
We observe that a typical meshlet is a connected patch of 2-manifold triangles, where  $\Vcur<\T<2\Vcur$ holds.
Thus, for configurations where a meshlet split algorithm is given limits of $\Tmax=\Vmax$, the limiting factor for the meshlet size is $\Tmax$, which is reached for most meshlets.
Vice versa, for configurations $\Tmax=2\Vmax$, the limiting factor for the meshlet size is $\Vmax$, which is reached for almost all meshlets, while the limit $\Tmax$ is rarely reached.
The workload per vertex, e.g., attribute transformations, is usually greater than the workload per triangle, e.g., loading three indices.
Therefore, to achieve higher \ac{GPU} utilization, it makes sense to set the limits such that $\Vcur$ is likely to reach a $\Vmax$ that is a multiple of the hardware's wave size.

As a meshlet index only references up to 256 unique vertices, the basic mesh-shading pipeline stores 8 bit per index, instead of 32 bits of a global index buffer of a vertex pipeline.
To achieve this, a meshlet stores a mesh-global offset to where its vertex attributes start in the buffer.
As a consequence, all the vertex attributes lying on the border between meshlets have to be duplicated, as each meshlet needs it in its local attribute space.
To reduce this data redundancy, meshlet builders can try to minimize the average border length of all meshlets.
It is possible to avoid attribute duplication entirely by using a secondary index buffer that references the attributes shared over multiple meshlets \cite{Kubisch18Introduction}.
While this would make the overall compression smaller, it makes the implementation more complex.
It also disrupts memory-locality of attributes, and thus makes memory access less coherent.
Further, it requires extra indirect memory accesses and makes streaming of individual meshlets more complex.
Finally, we want to utilize the geometric coherence between attributes of a meshlet to better quantize attributes in Sec. \ref{Sec:crack}.
Therefore, we prefer \emph{self-contained meshlets}, holding all attributes tightly in memory.
For our test cases, we found $\Vmax=128$ and $\Tmax=256$ to be optimal.
With $\Vmax=128$, 7 bit indices would suffice, but we choose 8-bit for faster byte-aligned memory access.

\section{Compression}
In this section, we formulate a \ac{MILP} for finding the optimal \ac{GTS}.
We then describe the strip encoding and parallel decoding.
We close this section with our crack-free attribute quantization.

\subsection{Optimal Generalized Triangle Strip}\label{Sec:OGTS}
To compress a meshlet's index buffer, we represent triangles as a \ac{GTS}.
For the best compression ratio, we need to find strips over the whole meshlet with the minimum possible number of restarts.
This problem is, however, NP-complete~\cite{Arkin96HTF} and many approximations exist, but see Vaněček's and Kolingerová's overview~\shortcite{Vanecek07CTS}.

Many practical optimization problems can be modeled as
\begin{equation}
\max \sum_i c_i v_i
\text{ such that } \mathbf{A}\vec{v}\preceq \vec{b}, v_i\geq 0,
\label{Eq:LPFormulation}
\end{equation}
with $\mathbf{A}\in\nreal^{m\times n}, \vec{b}\in\nreal^{m},\vec{c}\in\nreal^{n}$, $n$ the number of variables to be optimized, $m$ the number of constraints, and $\preceq$ the component-wise $\leq$ operator.
If $v_i\in\nreal$, Eq.~\ref{Eq:LPFormulation} is a \ac{LP}, if $v_i\in\nint$ an \acf{ILP}, and a \acf{MILP} when there is both.
Estkowski et al.~\shortcite{Estkowski02ODP} formulate \ac{ATS} computation as an \ac{ILP}.
We extend their approach to work for \acp{GTS}.
Consider the \emph{dual graph} of the triangle mesh.
A \emph{dual graph edge} connects two triangles, thus the \emph{dual graph nodes}, that share an edge.
Let $\vec{v}=\left[\vec{x}, \vec{y}\right]^\top$ in Eq.~\ref{Eq:LPFormulation}.
For each such edge $i$, the variable $x_i \in \cb{0;1}$ defines, whether it is part of the strip $(1)$ or not $(0)$.
The $x_i$ are the relevant solutions to our \ac{MILP}.
To minimize the required number of strip restarts, we need to maximize the number of edges that are part of the strip: $\max \sum_i x_i$.
As this condition alone does not create valid  triangle strips, we add two types of constraints:

\paragraph*{No-fork Constraint}
Since a manifold triangle has at most three neighboring triangles (one per edge) each dual-graph node has at most three dual-graph edges $x_b, x_l, x_r$.
If $x_b + x_l + x_r = 3$, all dual-graph edges would belong to the strip.
In such a case the strip would fork at this triangle, which is not possible.
Therefore, similar to Estkowski et al.~\shortcite{Estkowski02ODP}, we constrain
\begin{equation}
x_b + x_l + x_r\leq 2.
\end{equation}

\paragraph*{Anti-cycle Constraint}
To avoid unwanted cycles, Estkowski et al.~\shortcite{Estkowski02ODP} add one constraint per potential cyclic \ac{ATS}.
For \acp{ATS}, this is tangible because enumerating all cycles is $\mathcal{O}(n)$, but infeasible for \acp{GTS}.
Instead, solvers like Gurobi~\shortcite{Gurobi23GOL} use \emph{lazy constraints}, which can be dynamically added once a potential solution is found.
We observed that lazy anti-cycle constraints do not yield a solution within a reasonable time- or memory-frame.
\newcommand{\overbar}[1]{\mkern 2mu\overline{\mkern-2mu#1\mkern-2mu}\mkern 2mu}%
\newcommand{\edge}[1]{\overbar{#1}}%
As an alternative approach, Cohen~\shortcite{Cohen19SGP} derived anti-cycle constraints for common graph problems.
For finding a non-cyclic \ac{GTS}, we modify and simplify them to the following:
to each edge connecting some nodes $A$ and $B$, we assign two variables $y_{\edge{ab}}$ and $y_{\edge{ba}}$ on either side of the edge, where $y_i \in \nrealop$.
We constrain the sum of the two edge variables to an arbitrary constant value $F \in \nrealp$, e.g., $F=1$:
\begin{equation}\label{eq:flow}
y_{\edge{ab}} + y_{\edge{ba}} = F \cdot x_{\edge{ab}}.
\end{equation}
Thus, if the edge is part of the strip, so when $x_{\edge{ab}}=1$, the sum $y_{\edge{ab}} + y_{\edge{ba}}$ must be $F$.
Otherwise, if the edge is not part of the strip, both variables are zero.
Additionally, at each node, we constrain the sum of the adjacent edge flow variables, e.g., for a node $A$,
\begin{equation}\label{eq:flownode}
y_{\edge{ab}} + y_{\edge{ac}} + y_{\edge{ad}} < F.
\end{equation}
As strict inequality are not allowed in \acp{LP} programs, we write
\begin{equation*}
y_{\edge{ab}} + y_{\edge{ac}} + y_{\edge{ad}} \leq F - \epsilon
\end{equation*}
instead. Consider a strip section:
\newcommand{\halff}{\frac{F}{2}}
\newcommand{\edgelength}{1.8cm}
\begin{equation*}
\begin{tikzpicture}[
roundnode/.style={circle, draw=black!100, fill=white, very thick},
emptynode/.style={circle, draw=white, fill=white, very thick},
]
\node[emptynode] (X)              { };
\node[roundnode] (A) [right=of X] {$A$};
\node[roundnode] (B) [right=\edgelength of A] {$B$};
\node[roundnode] (C) [right=\edgelength of B] {$C$};
\node[emptynode] (Y) [right=of C] { };
\draw[loosely dotted, very thick] (X.east) -- (A.west);
\draw[-, very thick] (A.east) to node[very near start, above] {$y_{\edge{ab}}$} node[very near end, above] {$y_{\edge{ba}}$} (B.west);
\draw[-, very thick] (B.east) to node[very near start, above] {$y_{\edge{bc}}$} node[very near end, above] {$y_{\edge{cb}}$} (C.west);
\draw[loosely dotted, very thick] (C.east) -- (Y.west);
\end{tikzpicture},
\end{equation*}
which can be rewritten according to Eq.~\ref{eq:flow}:
\begin{equation*}
\begin{tikzpicture}[
roundnode/.style={circle, draw=black!100, fill=white, very thick},
emptynode/.style={circle, draw=white, fill=white, very thick},
]
\node[emptynode] (X)              { };
\node[roundnode] (A) [right=of X] {$A$};
\node[roundnode] (B) [right=\edgelength of A] {$B$};
\node[roundnode] (C) [right=\edgelength of B] {$C$};
\node[emptynode] (Y) [right=of C] { };
\draw[loosely dotted, very thick] (X.east) -- (A.west);
\draw[-, very thick] (A.east) to node[very near start, above] {$y_{\edge{ab}}$} node[near end, above] {$F - y_{\edge{ab}}$} (B.west);
\draw[-, very thick] (B.east) to node[very near start, above] {$y_{\edge{bc}}$} node[near end, above] {$F - y_{\edge{bc}}$} (C.west);
\draw[loosely dotted, very thick] (C.east) -- (Y.west);
\end{tikzpicture}.
\end{equation*}
With Eq.~\ref{eq:flownode}, it is given that $y_{\edge{ab}} > y_{\edge{bc}}$, or the other way round $y_{\edge{ba}} < y_{\edge{cb}}$.
As these inequalities cannot hold for a cyclic strip, we successfully prevent cycles with these additional constraints.

Finally, we feed our \ac{MILP} to the Gurobi~\shortcite{Gurobi23GOL} and SCIP~\shortcite{SCIP} solvers to find an optimal solution, as shown in our supplemental code.

\definecolor{teaserGreen}{RGB}{40, 173, 92}%
\definecolor{teaserRed}{RGB}{235, 59, 35}%
\begin{figure*}
	\centering
	\def\svgwidth{\textwidth}%
	\fontsize{8pt}{7pt}\selectfont%
	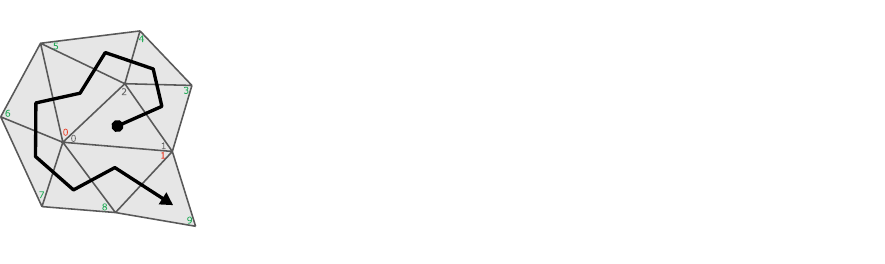
	\caption{Strip Encoding.
		\textbf{(a)} Example meshlet with vertex indices at the corners.
		The polyline denotes the triangle strip.
		\textbf{(b)} For \emph{Basic Mesh-Shading}, triangles use three indices stored in the \emph{Basic Index Buffer}.
		\textbf{(c)} Our meshlet \emph{\ac{GTS} Compression} consists of one index per triangle in an \emph{Index Buffer}. 
		The indices 0, 1, 2 are not explicitly stored.
		The \emph{L/R-Flag Buffer} stores a flag denoting the direction of the strip.		
		\textbf{(d)} Our meshlet \emph{\ac{GTS}-Reuse Compression} makes use of incrementing vertex indices ({\color{teaserGreen}green}) stored in an \emph{Increment-Flag Buffer}.
		In addition to the \emph{L/R-Flag Buffer}, the reused vertex indices ({\color{teaserRed}red}) have to be stored in the \emph{Reuse Buffer}.
		We compute \emph{Inclusively Add-Scanned Increment-Flag Buffer} during decompression.
		In case an index is reused, we compute the index location in the \emph{Reuse Buffer} with additions and subtractions.
	}\label{fig:Meshlets}
\end{figure*}

\subsection{Strip Encoding}\label{Sec:GTS}
As established in Sec.~\ref{sec:PreviousWork}, an L/R-flag per triangle marks the edge reused from the previous triangle.
As we decompress all triangles in parallel, a thread cannot access the decompressed previous triangle.
Instead, each thread must find the correct indices from earlier in the strip.
See Fig.~\ref{fig:Meshlets} for reference: the longer a \ac{GTS} spins around a vertex in a fan-like manner, the further back the required index lies:
here, triangle 8 requires the index of triangle 4, which is 0.

Meyer et al.~\shortcite{Meyer12DPD} use a max-scan to find this offset. 
Initial experiments with wave-intrinsics slowed our decompression.
Thus, we use \acp{GPU}-bit-level instructions on the L/R flags-for this search, similar to Nanite~\cite{karis2021nanite}:
With \emph{firstbithigh}~\cite{HLSLfirstbithigh}, we find the index of the first set bit of a word, starting at the most significant bit.
To make use of it, we shift and combine the flag words in a way that the most significant bit is the flag of the current triangle.
The next bit contains the flag of the triangle before the current triangle and so on.
In case the flag of the current triangle is set, we flip all bits before calling \emph{firstbithigh}, to search for the first unset bit.
For the rare case that the local fan exceeds 31 triangles, thus 32 bits of a word are not enough, we iteratively check previous words of the bit-flags.
To enable hardware triangle-back-face culling, we flip the triangle orientation depending on the L/R-flag.

Even our optimal \ac{GTS} algorithm may need more than one strip per meshlet.
Then, we require a strip \emph{restart}.
Using an individual restart bit per triangle would make the implementation more complex, and, as restarts are rare, they increase the memory footprint.
Similar to Meyer et al.~\shortcite{Meyer12DPD}, we use four degenerate triangles to emulate a restart.
In our evaluation, we show that degenerate triangles do not contribute much to the total run-time cost.
This increases the number of triangles in the strip.
In rare cases, we exceed the $\Tmax = 256$ limit and we must split the meshlet.

\subsection{Index Reuse Packing}\label{Sec:GTSReuse}
As the vertex attributes referenced by the index buffer are local per meshlet, we reorder them such that new vertices appear in the strip in ascending order.
This means a strip always starts with $\rb{0, 1, 2\dots}$.
Thus, the first triangle of a meshlet does not have to be stored explicitly.
Furthermore, the majority of indices are just increments of the previous index.
Meyer et al.~\shortcite{Meyer12DPD} make use of this redundancy by storing one bit per triangle that indicates whether the index appears before in the strip.
Indices which are increments of the previous index are marked with a $1$ increment flag.
Indices which are not increments, thus already appeared in the strip, are marked as reuse with a $0$ increment flag.
A prefix add scan over all flags, which is a common per-wave intrinsic, allows us to retrieved all incrementing indices.
If the flag of the current triangle $t$ is $1$, the result of the scan $s$ is the current index.
If the flag is $0$, an additional reuse array is accessed at location $t+1-s$, see Fig.~\ref{fig:Meshlets}.
To speed up computation, we again use a bit instruction over the increment flags and avoid synchronization between threads.
Thus, we count the number of set bits in a word, called \emph{countbits}~\cite{HLSLfirstbithigh}.

\subsection{Crack-Free Fixed-Point Vertex Attribute Quantization}\label{Sec:crack}
While developing our compression scheme, we observed that mesh shaders have a limited compute budget.
There, we have little room for sophisticated attribute compression.
As our goal is to beat the vertex pipeline, we settle with quantization.
We must, however, make sure that duplicate vertices along a meshlet boundary map to the same values.
Otherwise, we would get unwanted cracks.

The attributes of a vertex are organized as an \emph{attribute vector} $\vec{A} \in \nreal^n$.
A component $A_i$ is called \emph{attribute channel}.
For example, if a vertex consists of a position $\in \nreal^3$, a normal $\in \nreal^3$, and a texture coordinate $\in \nreal^2$, $\vec{A} \in \nreal^8$.
For convenient vertex-processing, attributes are usually floating-point quantized.
But the logarithmic quantization-curve of floating-point numbers does not concur with typical distributions found for attributes.
Therefore, it is memory-wasteful.
Instead, we use a simple uniform quantization with $b$ bits for each attribute channel.
We find $b=16$ bits for each attribute channel a sensible choice, because it allows fast memory-aligned attribute fetches and we found no visual deviations for our test meshes.
Further, the precision is higher than the 16-bit floating-point format commonly used in practice.

To avoid cracks, we propose to first map the meshlets to a global anisotropic grid with the following uniform spacings along each axis:
For each attribute channel $i$, such as the $x$ coordinate of the position, we find the meshlet with the largest extent $w_i$.
The sample spacing of the global grid for channel $i$ is then $\Delta_i = w_i/\left(2^b-1\right)$.
In general, the global grid requires more than $b$ bits for channel and is therefore more precise than the local grid.
For each meshlet, we store its lowest value $L_i$ for each channel $i$ with respect to the global grid.
The values of the attribute channel are stored relative to $L_i$. 
This guarantees that $b$ bits are sufficient for each attribute channel.
For decompression, an integer addition reconstitutes the attribute values to the global grid.
Then we map the values back to their original floating-point representation, used for standard vertex-processing.
This scheme is simple and suits the tight compute budget of mesh-shaders.

Assuming independent and identically distributed attributes, we get a greater or equal information content for the attribute channel $i$ than $b$.
Let the global mesh extent of the attribute channel $i$ be $W_i$.
The information content of this channel is then $\mathrm{log}_2\left(W_i/\Delta_i\right) \ge b$ bits for this channel, but see Tab.~\ref{tab:quantization} for concrete values.

\section{Results and Discussion}\label{Sec:Evaluation}
\begin{figure*}
\centering%
\def\svgwidth{\textwidth}%
\fontsize{6pt}{5pt}\selectfont%
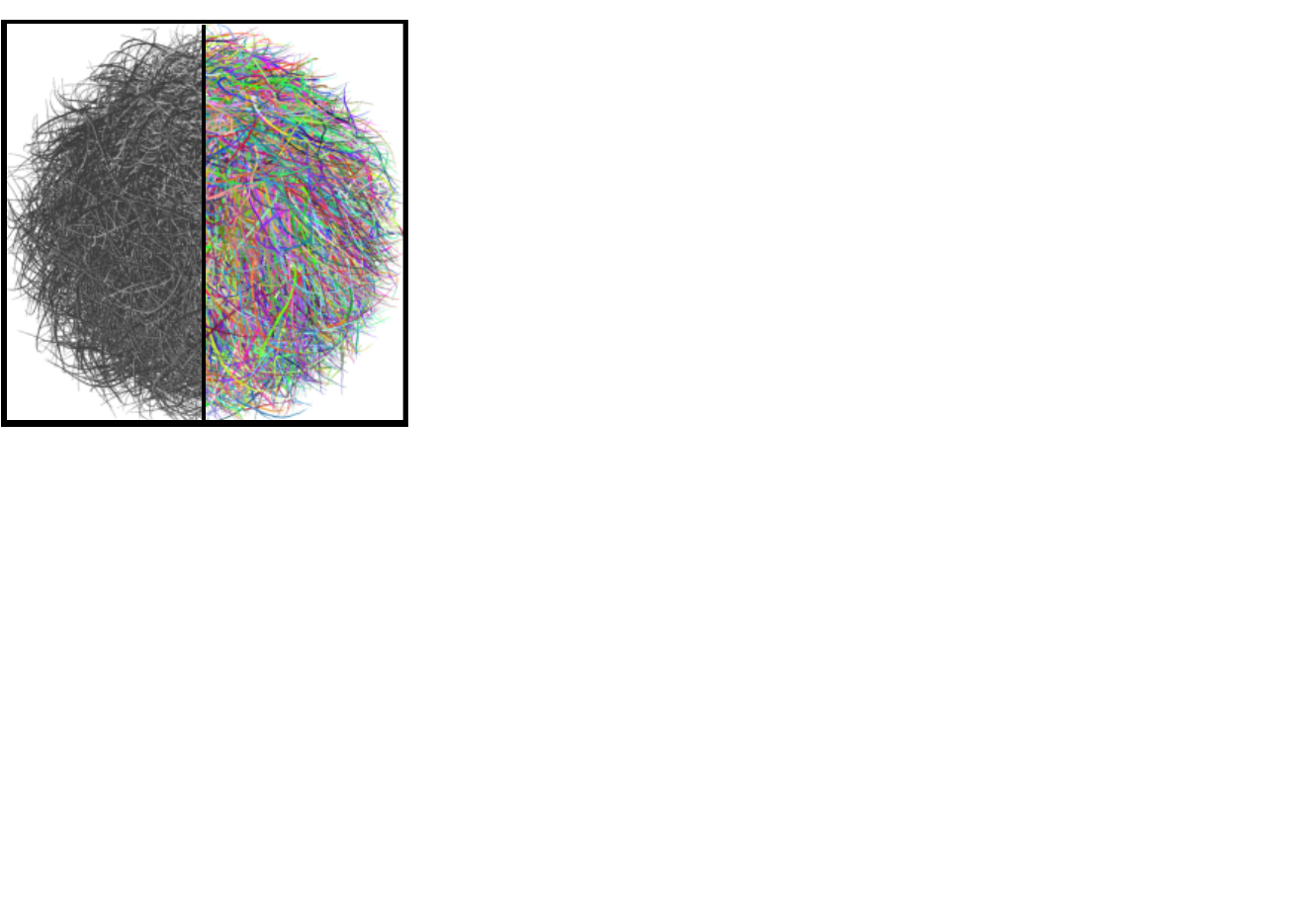%
\caption{Test Meshes.
$V$ denotes the number of vertices of the input mesh, $T$ the number of triangles and $M$ the number of meshlets coming from Meshoptimizer.
Meshlets are visualized with randomized colors.
}\label{fig:meshes}%
\end{figure*}%

We evaluate the performance of our methods on the meshes of Fig.~\ref{fig:meshes}.
Except for \textit{Hairball} and \textit{Cat}, our meshes come from scanning procedures.
\textit{Hairball} demonstrates that the graphics pipeline can be bound by other factors such as overdraw.
Meshlets were generated using the Meshoptimizer library~\cite{MeshOptimizer} with $\Vmax=128$ and $\Tmax=256$.
For the vertex pipeline experiments, we use vertex-cache and vertex fetch-optimization from Meshoptimizer.

\begin{table}
\centering
\footnotesize
\begin{tblr}[b]{lccc}
    \toprule                                        
                     & \textbf{Gurobi} & \textbf{SCIP}       & \textbf{ETA}\\
\midrule                                    
\textbf{computation time}     & 915.46\,s    & 8,122.56\,s & 10.52\,s   \\
\textbf{\ac{GTS} render time} & 0.58\,ms     & 0.58\,ms    & 0.59\,ms \\
\textbf{\ac{GTS} restarts}    & 123,784      & 123,784     & 185,480\\
\textbf{degenerate triangles} & 495,136      & 495,136     & 741,920\\
\textbf{additional meshlets}  & 3 / 75,412   & 3 / 75,412  & 4 / 75,413\\
\bottomrule                                    
\end{tblr}

%
\caption{\ac{GTS} comparison for the \textit{Rock} mesh.
We compare the optimal Gurobi and SCIP solutions against the sub-optimal \acs{ETA}.
The CPU computation uses one thread per meshlet and was measured on an AMD Ryzen 9 7950X (16C/32T).
}
\label{tab:stripify}
\end{table}

Tab.~\ref{tab:stripify} compares our optimal \acp{GTS} achieved with our \ac{MILP} of Sec.~\ref{Sec:OGTS} with Gurobi and SCIP against the sub-optimal strip of our \ac{ETA}~\cite{porcu2006partitioning} implementation.
As expected, the \ac{ETA} is orders of magnitude faster, but yields results worse than the globally optimal solution provided by the \ac{MILP} solvers.
Although \ac{ETA} requires about $1.5$ times as many strip restarts, and thus degenerate triangles, the difference in render time is negligible.
This also confirms the assumption that degenerate triangles do not contribute much to the run-time cost.

\begin{table*}
	\centering
	\footnotesize
\begin{tabular}[b]{l c c c c c c c c c c c}
\toprule     
\textbf{Mesh} 
& \multicolumn{4}{c}{\textbf{Index Buffer}}                                                   
& \multicolumn{2}{c}{\textbf{Meshlet Buffer}}
& \multicolumn{4}{c}{\textbf{Vertex Buffer}}\\
\cmidrule(lr){1-1}
\cmidrule(lr){2-5}
\cmidrule(lr){6-7}
\cmidrule(lr){8-11}
     & \multirow{2}{*}[-0.3em]{\makecell{Vertex \\ Pipeline}}
     & \multirow{2}{*}[-0.3em]{\makecell{Basic Mesh\\ Shading}} 
     & \multirow{2}{*}[-0.3em]{GTS}
     & \multirow{2}{*}[-0.3em]{\makecell{GTS\\ Reuse}}   
     & \multirow{2}{*}[-0.3em]{\makecell{Basic Mesh\\ Shading}}      
     & \multirow{2}{*}[-0.3em]{\makecell{GTS \& \\ GTS Reuse}}
     & \multirow{2}{*}[-0.3em]{\makecell{Vertex \\ Pipeline}}
     & \multirow{2}{*}[-0.3em]{\makecell{Basic Mesh\\ Shading}} 
     & \multicolumn{2}{c}{GTS \& GTS Reuse} \\     
     \cmidrule(lr){10-11}
     &  &     &  &                         &  &    &     &      & Float     & Fixed \\     
     \cmidrule(lr){2-5}
     \cmidrule(lr){6-7}
     \cmidrule(lr){8-11}
     \textbf{Hairball} & 32.6 (96) & 8.15 (24) & 3.21 (9.4) & 2.11 (6.2) &  0.328 & 0.336 &  44.6 & 48.0 & 48.1 & 25.3 \\
\textbf{Cat}      & 53.5 (96) & 13.4 (24) & 5.33 (9.6) & 3.32 (5.9) &  0.605 & 0.605 &  72.4 & 88.5 & 88.5 & 46.4 \\
\textbf{Skull}    & 57.7 (96) & 14.4 (24) & 5.71 (9.5) & 3.55 (5.9) &  0.651 & 0.651 &  77.7	& 95.2 & 95.2 & 49.9 \\
\textbf{Dragon}   & 82.6 (96) & 20.7 (24) & 8.17 (9.5) & 5.07 (5.9) &  0.931 & 0.931 &  110	& 136  & 136  & 71.4 \\
\textbf{Rock}     & 178  (96) & 44.5 (24) & 17.7 (9.5) & 11.0 (5.9) &  2.01 & 2.01  &  239  & 295  & 295  & 154  \\
\textbf{Lucy}     & 321  (96) & 80.3 (24) & 31.8 (9.5) & 19.7 (5.9) &  3.62 & 3.62  &  428  & 529  & 529  & 278	 \\
\bottomrule
\end{tabular}%
	\caption{Memory Requirements.
	We compare the sizes in MiB of the \emph{Index}, \emph{Meshlet}, and \emph{Vertex} buffers for the conventional \emph{Vertex Pipeline} and our mesh-shading pipelines (\emph{Basic Mesh Shading}, \emph{\ac{GTS}}, \emph{\ac{GTS}-Reuse}) for the respective \emph{Mesh}.
	The values in parentheses are \ac{bpt}. 
	Column \emph{Vertex Buffer} lists the memory consumption when using $8 \times 32$ bit \emph{Float}ing-point values per-vertex, except for column \emph{Fixed}, where attributes have $8 \times 16$ bit fixed-point values.
	\emph{Fixed} also includes the per-meshlet constants required for dequantization.}\label{tab:datasizes}
\end{table*}

Tab.~\ref{tab:datasizes} compares the memory requirements for different compression scenarios.
For the \emph{Index Buffer}, the basic mesh shader already achieves a compression ratio of $\sim\!\!4\!:\!1$ over the vertex pipeline.
With our \ac{GTS} compression of Sec.~\ref{Sec:GTS}, we achieve $\sim\!\!10\!:\!1$
By further packing the triangles with \ac{GTS}-Reuse of Sec.~\ref{Sec:GTSReuse}, we achieve $\sim\!\!16\!:\!1$.
With mesh-shaders, we store \emph{Meshlet Meta Buffer} such as buffer offsets (12 B) and cull information (16 B).
Due to degenerate triangles encoding restarts, both \ac{GTS} variants require extra meshlets in rare cases, where the triangle count exceeds $\Tmax$.

For \emph{Vertex Buffer}, we use eight attributes.
To leverage the advantages of self-contained meshlets described in Sec.~\ref{Sec:MeshShader}, we duplicate $\sim\!20\%$ of the vertices.
As expected, \emph{Fixed}-point quantization reduces the memory consumption by a factor of $\sim\!2\%$ over \emph{Float}ing-point quantization.
Tab.~\ref{tab:quantization} shows that our crack-free global-quantization increases the precision for the positions.
On the other hand, no precision is added to normal vectors, as a single meshlet with normal vectors cluttered to all directions is enough to mitigate the benefit.

\begin{table}
\centering%

\footnotesize
\begin{tabular}{cccccccc}
    \toprule    
\multicolumn{3}{c}{\textbf{Positions}} & \multicolumn{3}{c}{\textbf{Normals}} & \multicolumn{2}{c}{\textbf{Texture Coords.}} \\
                               \cmidrule(lr){1-3} \cmidrule(lr){4-6} \cmidrule(lr){7-8}
                    X   & Y    & Z                    & X     & Y    & Z      & U    & V                                \\
                    \cmidrule(lr){1-3} \cmidrule(lr){4-6} \cmidrule(lr){7-8}
      17.6 & 17.6 & 17.1                 & 16.0  & 16.0 & 16.0   & 16.1 & 16.0                             \\      
    \bottomrule
\end{tabular}%
\caption{Information content of our crack-free global-quantization on the \textit{Rock} mesh. 
We test each attribute channel of our meshlets with a fixed number of bits $b$, here $b=16$.
This allows for fast memory-aligned access.
Mapping the resulting sample-spacings from the meshlet-local grid to the global grid results in the shown global information content for the attribute values.
}%
\label{tab:quantization}%
\end{table}%

\begin{figure}
\centering
\begin{subfigure}[t]{\linewidth}%
\centering%
\def\svgwidth{\textwidth}%
\fontsize{4.5pt}{2pt}\selectfont%
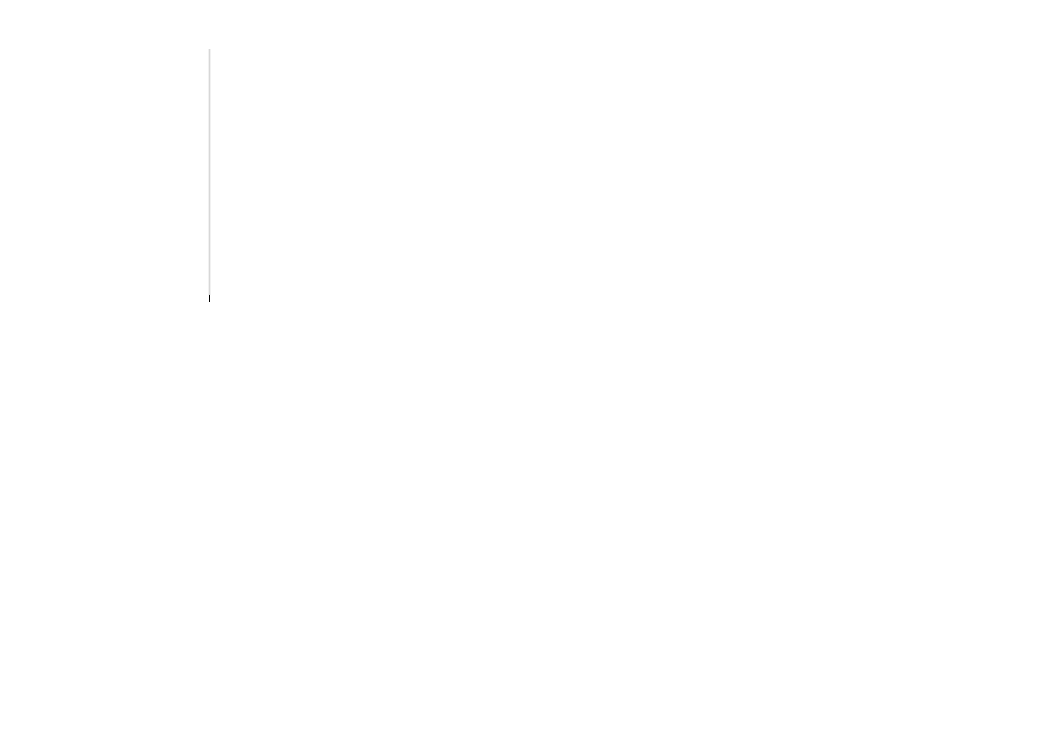%
\caption{Render Time without Cone-Culling.
}\label{fig:performance_timing}%
\end{subfigure}%
\newline%
\begin{subfigure}[t]{\linewidth}%
\centering%
\def\svgwidth{\textwidth}%
\fontsize{4.5pt}{2pt}\selectfont%
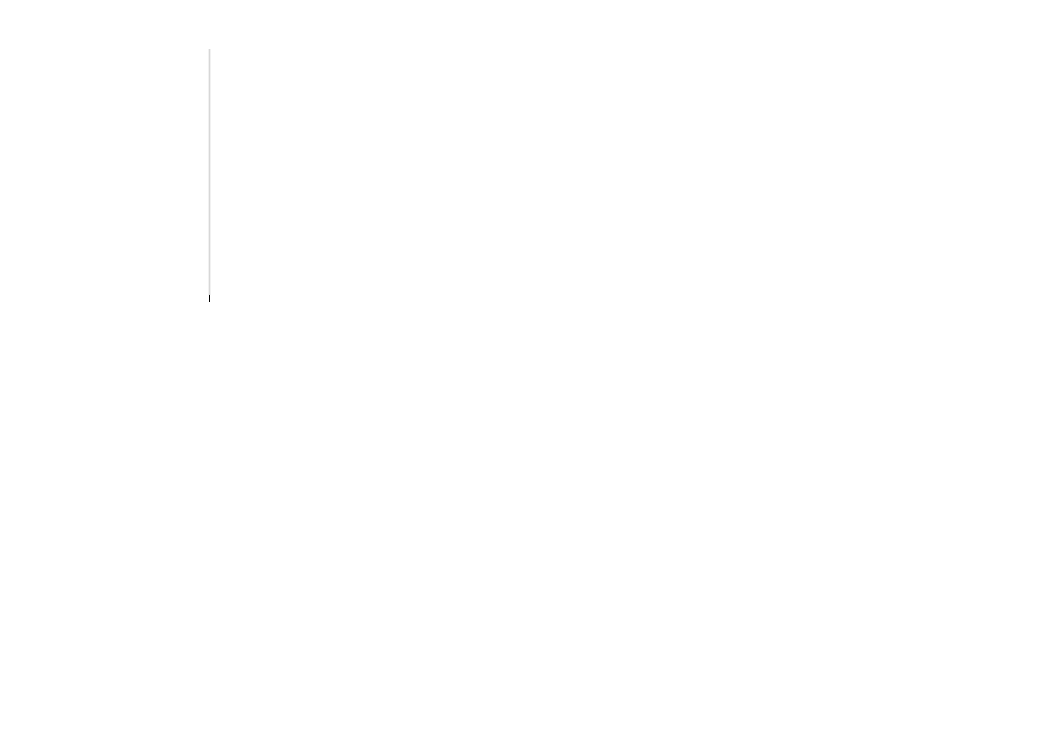%
\caption{
Render Time with Cone-Culling.
}\label{fig:performance_timing_culling}%
\end{subfigure}%
\newline%
\begin{subfigure}[t]{\linewidth}%
\centering%
\def\svgwidth{\textwidth}%
\fontsize{4.5pt}{2pt}\selectfont%
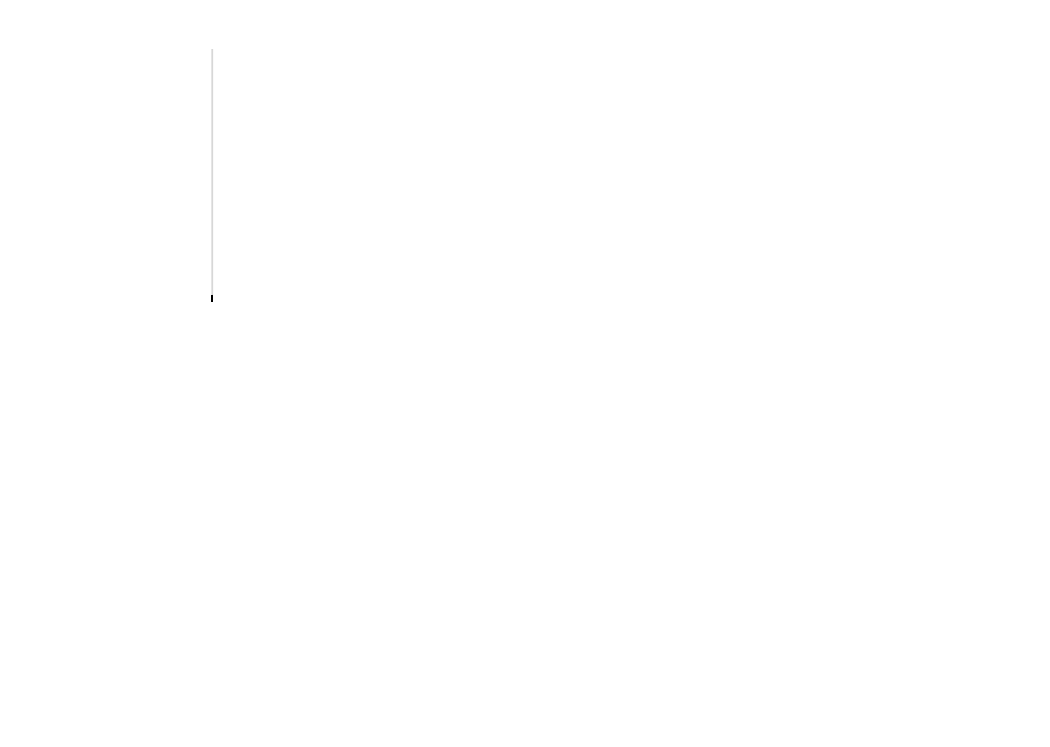%
\caption{Triangle Throughput without Cone-Culling.
}\label{fig:performance_throughput}%
\end{subfigure}%
\caption{
We compare the performance of different versions of our compression with (a) and without (b) cone-culling against the vertex pipeline (lower is better).
(c) To normalize for different mesh sizes, we compare the triangle throughput per second of our compression against the vertex pipeline (higher is better).
}\label{fig:performance}%
\end{figure}

Fig.~\ref{fig:performance} shows our rendering performance measurements on different \acp{GPU}.
To emphasize the performance impact of the geometry stage, our Direct3D12 implementation renders to a 500 by 500 pixel framebuffer with texture mapping and Phong Shading.
For more complex shading, the performance impact of the geometry stage on the total render-time decreases.
As expected, the render-time grows linearly with the mesh size.

To normalize different mesh sizes, Fig.~\ref{fig:performance_throughput} compares the triangle throughput per second.
Our basic mesh-shading pipeline outperforms the vertex pipeline.
When adding decompression, rendering is still faster than the vertex pipeline.
As expected, this is not always the case for the hairball, which is bound by pixel overdraw.

Fig.~\ref{fig:performance_timing_culling} is the same measurement as Fig.~\ref{fig:performance_timing}, but with cone culling enabled.
At most every second meshlet faces away from the camera, but in practise, this number is lower.
Therefore, we observe that cone culling improves performance by only $1.6$ times at most.
In the extreme case of our rock mesh, almost all meshlets face forward and cone-culling computations are in vain.

With our index and vertex buffer compression, the rendering performance is faster than the original vertex pipeline, when the pipeline is not bound by overdraw, but the memory footprint is significantly smaller.
More compression would only degrade performance, which counteracts our goal of beating the vertex pipeline in speed and size.
We observe that the compute capabilities of a mesh-shader are limited.

We compare against the meshlet compression scheme "\ac{LW}"~\cite{Mlakar24EEC}.
For connectivity, \ac{LW} requires 16~\ac{bpt}. 
With ca.\ 5.9~\ac{bpt}, our method is almost three times smaller.
\ac{LW} only reports numbers for positions, while we support an arbitrary number of attributes.
Therefore, we average index-, position-, and meshlet-buffer sizes over all models and obtain 38~\ac{bpt} for our method.
\ac{LW} compress on  their test corpus at an average of 37.5~\ac{bpt}.
Note that \ac{LW} quantizes positions to an average of $ 3 \times 15 = 45$~\ac{bpv}, whereas our approach is more precise with $3\times16=48$~\ac{bpv}.
When configuring our quantization algorithm to achieve \ac{LW}'s precision of $45$ \ac{bpv} for the Rock mesh, we need $13.4$, $13.4$, and $13.9$ bits for the three position components, totaling $40.7$~\ac{bpv}.
Due to better memory-alignment, we prefer power-of-two quantization.
\ac{LW} encoding timings range from "tens of minutes to hours for large scenes."
We encode the Rock mesh in $24$ s (\ac{ETA}) and $15$ minutes (Gurobi) including meshlet generation and quantization.
On an Nvidia RTX 4090, \ac{LW} decompresses and renders at 13.3~\ac{Gtps} \emph{with frustum and cone culling} optimization, only three attributes, and no textures.
Our slowest approach \emph{without meshlet culling}, with eight attributes, and textures achieves 22.7 -- 26.7~\ac{Gtps} and up to 36.8~\ac{Gtps} with cone culling.
To conclude the comparision, without culling, our approach is twice as fast, with cone-culling three times faster, it is more precise, and achieves comparable compression ratios.

\section{Conclusion and Future Work}\label{Sec:ConclusionAndFutureWork}
We proposed meshlet codecs.
To compress the topology of a meshlet, we find the optimal \ac{GTS}.
Index coherence within the strip is then used for further compression.
Our method compresses the input index buffer of the conventional vertex pipeline at a ratio of up to 16:1.
Furthermore, we demonstrated how to perform memory-aligned and crack-free attribute quantization, while making use of the limited local value range within a meshlet.
Our evaluation shows that our decompression runs faster than the vertex pipeline.
As future work we see specialized meshlet builders and more compact attribute representations.
We include code for \acs{GPU} decoding, quantization, and our \ac{MILP} in the supplemental material.

\section*{Acknowledgments}
\ifx\anonymize\undefined%
We thank Dominik Baumeister.
\fi%
Meshes are courtesy of Aixterior (Rock, Skull), Morgan McGuire (Hairball), the Stanford 3D Scanning Repository (Dragon, Lucy), and Ekaterina Kozemerchak (Cat).
 
\bibliographystyle{eg-alpha-doi}
\bibliography{main}     

\newcommand{\etalchar}[1]{$^{#1}$}
\begin{thebibliography}{\uppercase{MLDH15}}

\bibitem[AG03]{Alliez03RAC}
\textsc{Alliez P., Gotsman C.}:
\newblock Recent advances in compression of {3D} meshes.
\newblock In \emph{Advances in Multiresolution for Geometric Modelling} (2003),
  Springer Berlin Heidelberg.

\bibitem[AHMS96]{Arkin96HTF}
\textsc{Arkin E.~M., Held M., Mitchell J. S.~B., Skiena S.~S.}:
\newblock Hamiltonian triangulations for fast rendering.
\newblock \emph{The Visual Computer 12}, 9 (1996), 429--444.

\bibitem[BBC{\etalchar{*}}21]{SCIP}
\textsc{Bestuzheva K., Besan{\c{c}}on M., Chen W.-K., Chmiela A., Donkiewicz
  T., van Doornmalen J., Eifler L., Gaul O., Gamrath G., Gleixner A., Gottwald
  L., Graczyk C., Halbig K., Hoen A., Hojny C., van~der Hulst R., Koch T.,
  L{\"u}bbecke M., Maher S.~J., Matter F., M{\"u}hmer E., M{\"u}ller B.,
  Pfetsch M.~E., Rehfeldt D., Schlein S., Schl{\"o}sser F., Serrano F., Shinano
  Y., Sofranac B., Turner M., Vigerske S., Wegscheider F., Wellner P., Weninger
  D., Witzig J.}:
\newblock \emph{{The SCIP Optimization Suite 8.0}}.
\newblock Technical report, Optimization Online, December 2021.
\newblock URL:
  \url{http://www.optimization-online.org/DB_HTML/2021/12/8728.html}.

\bibitem[Cal02]{Calver2002VDS}
\textsc{Calver D.}:
\newblock \emph{Vertex and Pixel Shader Tips and Tricks}.
\newblock Wordware Publishing, 2002, ch.~Vertex Decompression in a Shader,
  pp.~172 -- 187.

\bibitem[CDE{\etalchar{*}}14]{Cigolle2014SER}
\textsc{Cigolle Z.~H., Donow S., Evangelakos D., Mara M., McGuire M., Meyer
  Q.}:
\newblock A survey of efficient representations for independent unit vectors.
\newblock \emph{Journal of Computer Graphics Techniques (JCGT) 3}, 2 (April
  2014).

\bibitem[Coh19]{Cohen19SGP}
\textsc{Cohen N.}:
\newblock {Several Graph problems and their Linear Program formulations}.
\newblock working paper or preprint, Jan. 2019.
\newblock URL: \url{https://hal.inria.fr/inria-00504914}.

\bibitem[Dee95]{Deering95GC}
\textsc{Deering M.}:
\newblock Geometry compression.
\newblock SIGGRAPH '95.

\bibitem[EMX02]{Estkowski02ODP}
\textsc{Estkowski R., Mitchell J. S.~B., Xiang X.}:
\newblock Optimal decomposition of polygonal models into triangle strips.
\newblock In \emph{Proceedings of the Eighteenth Annual Symposium on
  Computational Geometry} (New York, NY, USA, 2002), SCG '02, Association for
  Computing Machinery, p.~254–263.

\bibitem[FH11]{Frey11SSD}
\textsc{Frey I.~Z., Herzeg I.}:
\newblock Spherical skinning with dual quaternions and qtangents.
\newblock In \emph{ACM SIGGRAPH 2011 Talks} (2011), SIGGRAPH '11, Association
  for Computing Machinery.

\bibitem[GLLR11]{Gurung11LRC}
\textsc{Gurung T., Luffel M., Lindstrom P., Rossignac J.}:
\newblock Lr: compact connectivity representation for triangle meshes.
\newblock \emph{ACM Trans. Graph. 30}, 4 (2011).

\bibitem[GS98]{Gumhold98CBM}
\textsc{Gumhold S., Stra\ss{}er W.}:
\newblock Real time compression of triangle mesh connectivity.
\newblock In \emph{Proceedings of the 25th Annual Conference on Computer
  Graphics and Interactive Techniques} (1998), Association for Computing
  Machinery.

\bibitem[{Gur}23]{Gurobi23GOL}
\textsc{{Gurobi Optimization, LLC}}:
\newblock {Gurobi Optimizer Reference Manual}, 2023.
\newblock URL: \url{https://www.gurobi.com}.

\bibitem[JBG17]{Jakob17Pac}
\textsc{Jakob J., Buchenau C., Guthe M.}:
\newblock A parallel approach to compression and decompression of triangle
  meshes using the {GPU}.
\newblock \emph{Comput. Graph. Forum 36}, 5 (Aug. 2017).

\bibitem[Kap23]{MeshOptimizer}
\textsc{Kapoulkine A.}:
\newblock {meshoptimizer}, 2023.
\newblock URL: \url{https://github.com/zeux/meshoptimizer}.

\bibitem[Kil08]{Kilgard08MOU}
\textsc{Kilgard M.}:
\newblock \emph{{Modern OpenGL usage: Using vertex buffer objects well}}.
\newblock Tech. rep., NVIDIA Corporation, 2008.

\bibitem[KISS15]{Keinert15SFM}
\textsc{Keinert B., Innmann M., S\"{a}nger M., Stamminger M.}:
\newblock Spherical fibonacci mapping.
\newblock \emph{ACM Trans. Graph. 34}, 6 (Oct. 2015).

\bibitem[KM21]{Kuth21VBA}
\textsc{Kuth B., Meyer Q.}:
\newblock Vertex-blend attribute compression.
\newblock In \emph{High-Performance Graphics - Symposium Papers} (2021), Binder
  N., Ritschel T., (Eds.), The Eurographics Association.

\bibitem[KSW21]{karis2021nanite}
\textsc{Karis B., Stubbe R., Wihlidal G.}:
\newblock A deep dive into nanite virtualized geometry.
\newblock In \emph{ACM SIGGRAPH} (2021).

\bibitem[Kub18]{Kubisch18Introduction}
\textsc{Kubisch C.}:
\newblock \emph{Introduction to {Turing} Mesh Shaders}, 09 2018.
\newblock URL:
  \url{https://developer.nvidia.com/blog/introduction-turing-mesh-shaders/}.

\bibitem[KXW{\etalchar{*}}18]{Kwan18PVD}
\textsc{{Kwan} K.~C., {Xu} X., {Wan} L., {Wong} T., {Pang} W.}:
\newblock Packing vertex data into hardware-decompressible textures.
\newblock \emph{IEEE Transactions on Visualization and Computer Graphics 24}, 5
  (2018).

\bibitem[LCL10]{Lee10C3M}
\textsc{Lee J., Choe S., Lee S.}:
\newblock {Compression of 3D Mesh Geometry and Vertex Attributes for Mobile
  Graphics}.
\newblock \emph{JCSE 4} (09 2010).

\bibitem[Mey12]{Meyer12RTG}
\textsc{Meyer Q.}:
\newblock \emph{Real-Time Geometry Decompression on Graphics Hardware}.
\newblock PhD thesis, 08 2012.

\bibitem[Mic20]{HLSLfirstbithigh}
\textsc{Microsoft}:
\newblock \emph{{High-level shader language (HLSL)}}, 08 2020.
\newblock URL:
  \url{https://learn.microsoft.com/en-us/windows/win32/direct3dhlsl/dx-graphics-hlsl}.

\bibitem[Mic23]{DirectXSpecs}
\textsc{Microsoft}:
\newblock \emph{DirectX-Specs}, 2023.
\newblock URL: \url{https://microsoft.github.io/DirectX-Specs/}.

\bibitem[MKSS12]{Meyer12DPD}
\textsc{Meyer Q., Keinert B., Sußner G., Stamminger M.}:
\newblock Data-parallel decompression of triangle mesh topology.
\newblock \emph{Computer Graphics Forum 31}, 8 (2012), 2541--2553.

\bibitem[MLDH15]{Maglo153MC}
\textsc{Maglo A., Lavou\'{e} G., Dupont F., Hudelot C.}:
\newblock 3d mesh compression: Survey, comparisons, and emerging trends.
\newblock \emph{ACM Comput. Surv. 47}, 3 (2015).

\bibitem[MMT23]{Maggiordomo23MMC}
\textsc{Maggiordomo A., Moreton H., Tarini M.}:
\newblock Micro-mesh construction.
\newblock \emph{ACM Trans. Graph. 42}, 4 (2023).

\bibitem[MSGS11]{Meyer11ALP}
\textsc{Meyer Q., Su\ss{}ner G., Greiner G., Stamminger M.}:
\newblock Adaptive level-of-precision for {GPU}-rendering.
\newblock In \emph{Vision, Modeling, and Visualization (2011)} (2011), The
  Eurographics Association.

\bibitem[MSS{\etalchar{*}}10]{Meyer10OFN}
\textsc{Meyer Q., S\"{u}\ss{}muth J., Su\ss{}ner G., Stamminger M., Greiner
  G.}:
\newblock On floating-point normal vectors.
\newblock In \emph{Proceedings of the 21st Eurographics Conference on
  Rendering} (2010), EGSR'10, Eurographics Association.

\bibitem[MSS24]{Mlakar24EEC}
\textsc{Mlakar D., Steinberger M., Schmalstieg D.}:
\newblock End-to-end compressed meshlet rendering.
\newblock \emph{Computer Graphics Forum 43}, 1 (2024).

\bibitem[{Nvi}22]{Nvidia23MM}
\textsc{{Nvidia}}:
\newblock {Displacement-MicroMap-Toolkit}, 2022.
\newblock URL:
  \url{https://github.com/NVIDIAGameWorks/Displacement-MicroMap-Toolkit}.

\bibitem[PBCK05]{Purnomo05HCV}
\textsc{Purnomo B., Bilodeau J., Cohen J.~D., Kumar S.}:
\newblock Hardware-compatible vertex compression using quantization and
  simplification.
\newblock HWWS '05, Association for Computing Machinery.

\bibitem[PKJ05]{Peng05T3M}
\textsc{Peng J., Kim C.-S., {Jay Kuo} C.-C.}:
\newblock {Technologies for 3D mesh compression: A survey}.
\newblock \emph{Journal of Visual Communication and Image Representation 16}, 6
  (2005).

\bibitem[PKM22]{Peters22PCV}
\textsc{Peters C., Kuth B., Meyer Q.}:
\newblock Permutation coding for vertex-blend attribute compression.
\newblock \emph{Proc. ACM Comput. Graph. Interact. Tech. 5}, 1 (may 2022).

\bibitem[PS06]{porcu2006partitioning}
\textsc{Porcu M.~B., Scateni R.}:
\newblock {Partitioning Meshes into Strips using the Enhanced Tunnelling
  Algorithm (ETA)}.
\newblock In \emph{VRIPHYS} (2006), pp.~61--70.

\bibitem[Ros99]{Rossignac99ECC}
\textsc{Rossignac J.}:
\newblock Edgebreaker: Connectivity compression for triangle meshes.
\newblock \emph{IEEE transactions on visualization and computer graphics 5}, 1
  (1999), 47--61.

\bibitem[SPM{\etalchar{*}}12]{Schaefer12MAT}
\textsc{Sch\"{a}fer H., Prus M., Meyer Q., S\"{u}\ss{}muth J., Stamminger M.}:
\newblock Multiresolution attributes for tessellated meshes.
\newblock In \emph{Proceedings of the ACM SIGGRAPH Symposium on Interactive 3D
  Graphics and Games} (2012), Association for Computing Machinery.

\bibitem[Tut62]{Tutte62CPT}
\textsc{Tutte W.~T.}:
\newblock A census of planar triangulations.
\newblock \emph{Canadian Journal of Mathematics 14} (1962), 21–38.

\bibitem[VdFG99]{Velho99HGT}
\textsc{Velho L., de~Figueiredo L.~H., Gomes J.}:
\newblock Hierarchical generalized triangle strips.
\newblock \emph{The Visual Computer 15}, 1 (1999), 21--35.

\bibitem[VK07]{Vanecek07CTS}
\textsc{Vaněček P., Kolingerová I.}:
\newblock Comparison of triangle strips algorithms.
\newblock \emph{Computers and Graphics 31}, 1 (2007), 100--118.

\end{thebibliography}

\end{document}